\documentstyle[11pt,newpasp,twoside,epsfig]{article}
\newcommand{\mcol}[3]{\multicolumn{#1}{#2}{#3} }

\newcommand{\struut}{\rule[-2ex]{0ex}{5.2ex}}
\newcommand{\struutup}{\rule{0ex}{3.2ex}}
\newcommand{\struutdown}{\rule[-2ex]{0ex}{2ex}}
\newcommand{\ap}{$\approx$}

\begin{document}

\title{$\delta$ Scuti Stars in Stellar Systems: a Review}

\author{P. Lampens and H.M.J. Boffin}
\affil{Royal Observatory of Belgium, Ringlaan 3, B-1180 Brussels, Belgium}

\begin{abstract}
We present a list of $\delta$ Scuti stars in double and multiple systems, ranging
from the very wide binaries to the very close ones such as spectroscopic and eclipsing 
systems including the optical visual pairs which are of no further use here. Our aim
is to group the information from the binarity on the one hand and the pulsational 
characteristics on the other hand for as complete a sample as possible 
of $\delta$ Scuti stars in stellar systems. A selection of 18 well-documented cases, 
taking care that every type of binary is being represented, is discussed more extensively.  
\end{abstract}


\section{Introduction}

It is well known that the vast majority of stars belong to a binary
or a multiple system, irrespective of their spectral type. Recent surveys with 
improved astrometric accuracy, either from space or from the ground, present 
clear evidence that the higher the accuracy, the larger the number of binary 
detections. Up to 3000 new binaries have thus been discovered during the Hipparcos 
satellite mission (Lindegren 1997). The frequency of binaries is estimated to be
at least 60\% in the solar neighbourhood (Duquennoy \& Mayor 1991) but this
is probably an underestimation as modelisation tends to show. For example,
Odenkirchen and Brosche (1999) found that a frequency of at least 70\%
was needed in some models to account for the existence of another 2400
(up to now undetected) astrometric binaries in the Hipparcos catalogue.
The same is also true with respect to the improvement of the photometric
accuracy: the higher the accuracy, the larger the number of variable star 
detections. The results of the Hipparcos photometric survey are again 
quite illustrative: some 8000 new variable stars have been identified 
(van Leeuwen 1997). It is thus not at all a surprise to encounter pulsating
$\delta$ Scuti stars as components of stellar systems. The catalogue of Seeds 
and Yanchak (1972) contains a third of double stars over a total of 155 $\delta$ 
Sct variables or suspected ones. Szatm\'ary (1990) also quotes a binary fraction 
of about 30 \%. 
To study such cases presents some obvious advantages: we have additional constraints 
on some important physical parameters such as distance, mass, radius... but this of 
course will depend on the type of binary or multiple system of which the pulsating
star is a member. We may find them among the wide visual binaries (VB), the closer
visual binaries with orbital motion (VB/O), the (unresolved) astrometric binaries (AB),
the spectroscopic binaries (SB), or even eclipsing and ellipsoidal binary systems (E).
In almost all these cases it is generally accepted that both components pertaining 
to the system originate from the same parent cloud, therefore have the same
age and chemical composition. Additional information on the non-variable component
can then be extrapolated unto the variable one and used to better constrain the
variability characteristics such as the pulsation type and modes.
But we could go further and address such questions as: does binarity modify
the pulsation properties or even trigger the pulsation? Is there any correlation 
between rotation, pulsation and orbital motion ?\\
For all these reasons the identification of such objects offers an interesting
challenge toward a better comprehension of pulsation in the $\delta$ Scuti 
instability strip. A comparative study between the two companions allows to 
restrict for example the domain in mass in which the $\delta$ Sct phenomenon takes 
place. Various recent studies exist that investigate the pulsating $\delta$ Scuti 
stars in open clusters for similar reasons (e.g. the Praesepe cluster (Pe\~na et al.~1998; 
Alvarez et al.~1998)). One should however also keep in mind that star formation in some 
open clusters is not always a single-epoch process (e.g. M16 (Hillenbrand et al. 1993)).

\section{Summary tables}

We here summarize the available information for as complete a sample as possible 
of $\delta$ Scuti stars in stellar systems in two tables, one for the binarity
and one for the variability characteristics. The basis was drawn on the catalogue 
work by Garc\'{\i}a et al. (1995), supplemented by our own list and literature search. 
The Catalogue of the Components of Double and Multiple Stars (CCDM, Dommanget \& Nys 1995) 
was used to retrieve information on the visual components of the stellar systems.
Table~1. lists the name and the HD number (or BD number if HD does not exist), the Hipparcos 
number, the V magnitude, remarks concerning the multiplicity, the Hipparcos parallax and error. 
Next comes the information regarding the duplicity in the Hipparcos Catalogue (ESA 1997): a flag 
referring to that part of the Annex of Double and Multiple Systems (DMSA) where the star 
has been classified, the component designations, the angular se\-pa\-ration (in arcsec), 
followed by the information about the variability: the periodicity, a flag indicating in
which Variability Annex the star has been classified (P/1=periodic variable; U/2=unsolved;
M=possible microvariable; D='duplicity-induced variability'; C='constant').
The final columns list the median Hp magnitude with standard error and the differential 
Hp magnitude with standard error.\\
Table~\ref{Tab:variability}~lists the name and the HD number (or BD number if HD does not exist),
remarks concerning the variability, the number of frequencies, the dominant periodicity
and semi-amplitude, the projected rotational velocity of the $\delta$ Sct star, 
the spectral types of the $\delta$ Sct and additional components. Among the remarks the 
following notations are used: N=narrow range of frequencies; W=wide range of frequencies; 
R=radial pulsator; NR=non-radial pulsator.

\begin{figure*}
\setlength{\textwidth}{17cm}
\plotone{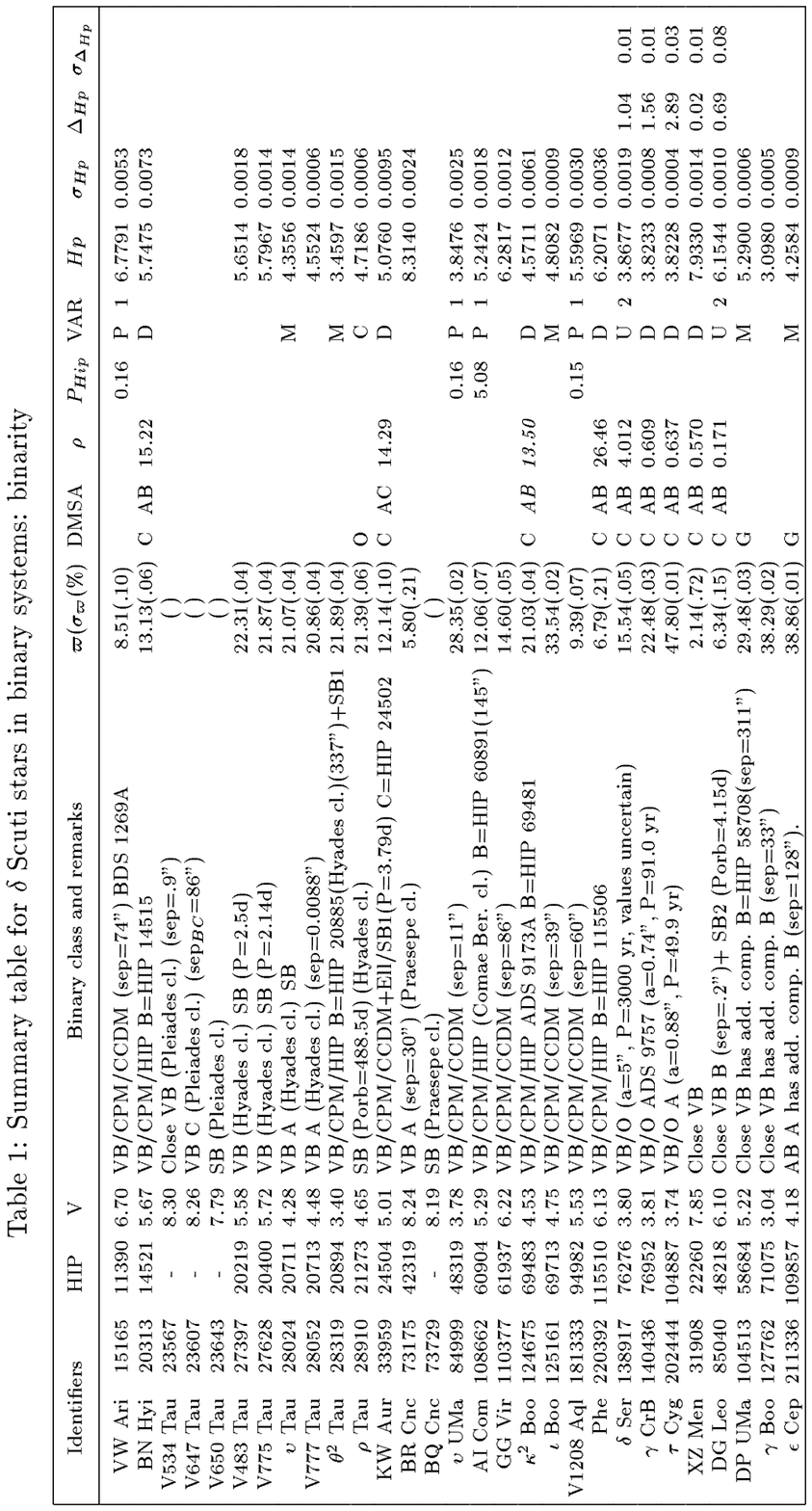}
\end{figure*}
\begin{figure*}
\setlength{\textwidth}{17cm}
\plotone{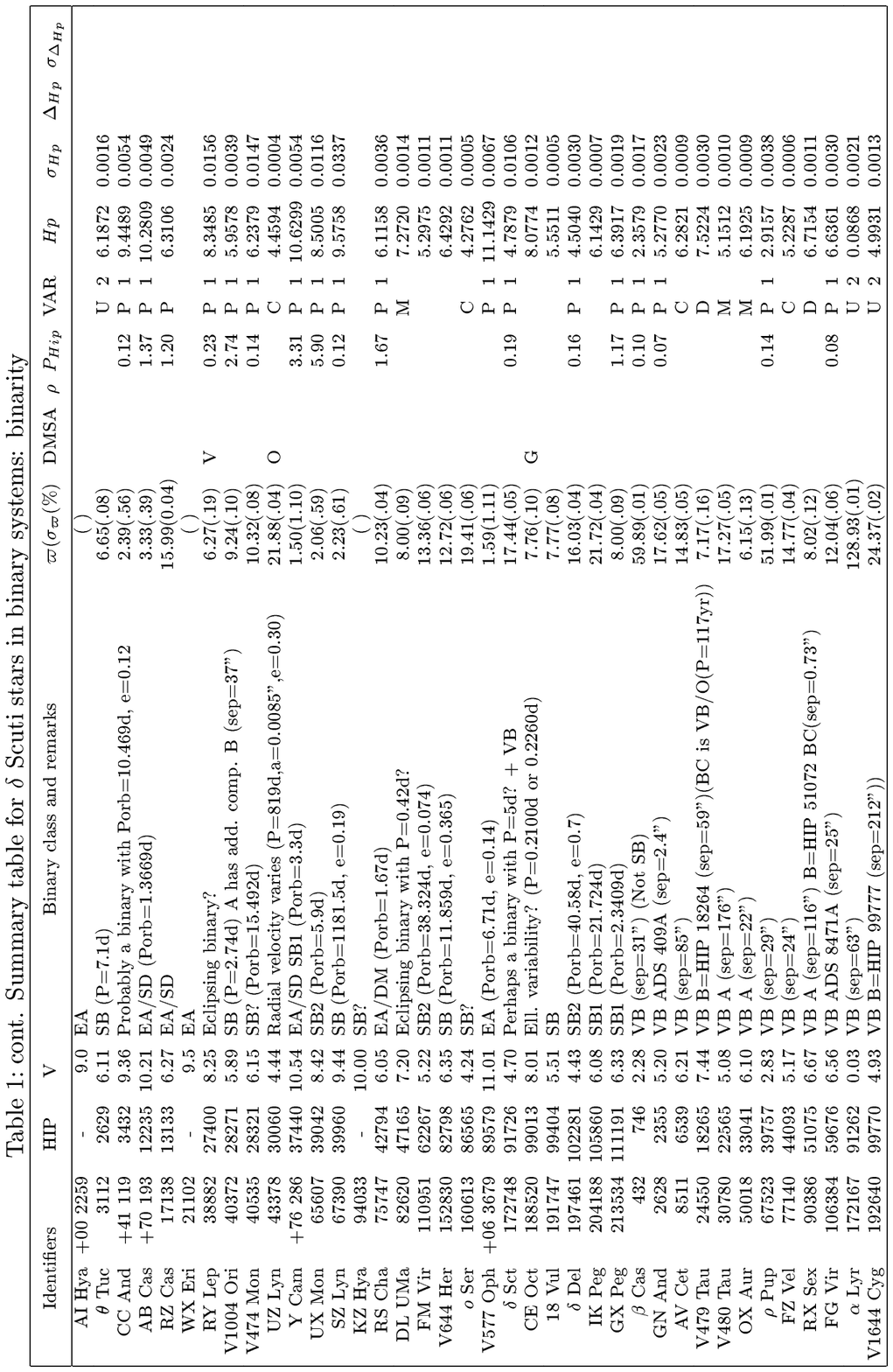}
\end{figure*}
\addtocounter{table}{1}

\begin{table} \label{Tab:variability}
\begin{center}
\setlength{\tabcolsep}{0.5mm}
\caption[]{Summary table for $\delta$ Scuti stars in binary systems: \\
\hspace*{5cm}variability \hfill}
{\tiny
\begin{tabular}{|rrcclllll|} 
\hline 
\mcol{2}{|c}{Identifiers}& 
\mcol{1}{c}{Pulsation}& \mcol{1}{c}{\#Freq.}&
\mcol{1}{c}{Period}& \mcol{1}{c}{Amp(V/B)}& 
\mcol{1}{c}{$v_{1}$sin{\it i}}& \mcol{1}{c}{$SpT_{1}$}& 
\mcol{1}{c|}{$SpT_{2}$ \struut}\\
\hline
VW       Ari&   15165& W 	    &$>7$     & P1=0.161 &A1=0.020&  90 &F0IV &B=F2IV \struutup\\
BN       Hyi&   20313&		    &  2      & 0.065	 & 0.006  &  47 &F2II-III    & -	  \\			 
V534     Tau&   23567&		    &	      & 0.032	 & 0.009  &  90 &A9V	     & -	  \\			 
V647     Tau&   23607&		    &	      & 0.047	 & 0.007  &  10 &A7V	     & -	  \\		 
V650     Tau&   23643 &R+NR;W	    &  5      & 0.031	 & 0.004v?& 185 & -	     & -	  \\		 
V483     Tau&   27397 &Q=0.038&      & 0.0548   & 0.013  & 100 &F3V...	     & -	  \\	       
V775     Tau&   27628&		    &	      & 0.06 &$\leq$ 0.005&  30 &A3m	     & -	  \\		 
$\upsilon$ Tau&   28024&		  &         & 0.148    & 0.010  & 195 &A8Vn	   & -  	\\	       
V777     Tau&   28052&		  &         & 0.182    & 0.010  & 195 &F0V...	   & -  	\\       
$\theta ^2$ Tau&   28319&N 	  &$>5$&P1=0.0756 &A1=0.007& 140 &A7III       & B=G7III    \\
$\rho$   Tau&   28910&		    &	      & 0.07	 & 0.005  & 103 &A8V	     & -      \\				
KW       Aur&   33959&Q=0.012;N      &3&P1=0.0881 &A1=0.016&  25 &A9IV  & C=F3V    \\
BR       Cnc&   73175&		    &$>$3     & 0.040	 & 0.010  & 170 &F0Vn	     & -      \\     
BQ       Cnc&   73729&               &  3      & 0.064	   & 0.003  & 165 &F2V         & -	    \\	   
$\upsilon$ UMa&   84999&NR high {\it m}&many  & 0.07-0.09&\ap0.03 & 115 &F0IV        & -	    \\	   
AI       Com&  108662&		    &  1?     & 0.052	 & 0.02   &  10 &A0p	     &A0p     \\  
GG       Vir&  110377&		    &many&  0.023$\leq$P$\leq$0.083& $\leq$0.014& 160 &A7Vn	   &G8V       \\	     
$\kappa ^2$ Boo&  124675&Q=0.016;N  &4&  0.0648  & 0.0064 & 140 &A8IV    &F1V	    \\		    
$\iota$  Boo&  125161&susp.$\delta$ Sct& 1?  & 0.026    & 0.0035 & 130 &A9V         & -	    \\			  
V1208    Aql&  181333& Q=0.034;N	  &$>2$     &P1=0.146v&A1=0.013v&  48 &F0III       &G5	    \\ 
   -     Phe&  220392& N		  &  2      &P1=0.2139 &A1=0.014& 145 &F0IV        &B=A9V	    \\     
$\delta$ Ser&  138917&   		  & 2 & 0.156&0.01-0.02& 70  &F0IV        & -	    \\		       
$\gamma$ CrB&  140436&		  &$>2$ &0.111or0.296 &K=1-2 kms$^{-1}$& 100 &A1Vs        & -	    \\ 	    
$\tau$   Cyg&  202444&  	     &       &0.083$\leq$P$\leq$0.125&0.01&  91 &F1IV	     & -       \\ 
XZ       Men&   31908&  	     &many     &P1=0.1083 &A1=0.009&  -  &Fm...       & -      \\      
DG       Leo&   85040&  	     &$>1$?    &0.0826   & 0.015  &  18 &A8IV	     & Am?   \\		  
DP       UMa&  104513&  	     &many     & 0.04var&$\leq$0.01&  78 &A7m	      &B=G8V	       \\      
$\gamma$ Boo&  127762&NR high {\it m}&2&0.073&0.050& 135 &A7IIIvar    & -	    \\  				   
WX       Eri&   21102&susp. $\delta$ Sct   &    & 0.1645   & 0.010  &  -  &A7          & -	    \\
$\epsilon$ Cep&  211336&NR high {\it m}&$>2$? &P1=0.041  &A1=0.008& 105 &F0IV        & -	    \\ 
AI       Hya&+00 2259 &               &    &     0.1380	& 0.020  &  24  &F2	    & - 	 \\
$\theta$ Tuc&    3112&unstable?;W  & 10&P=0.049   &0.013 &  80   &A7IV        & -	   \\ 			
CC       And&   +41 119~    &  	     &  7 &	0.1249  &0.24  &  20	 &F3IV-V      & -      \\		  
AB       Cas&   +70 193~    &  Q=0.036      &  1 &	0.0583 &0.03$\leq$A$\leq$0.09v&  55 &A3V    &K1V	 \\		 
RZ 	 Cas&	17138& susp. $\delta$ Sct&many& 0.016	& 0.02& -  &A3V & K0IV \\
RY       Lep&   38882&  	     &    &	0.2254   & 0.35 	&  -  &A9V   & -       \\			  
V1004    Ori&   40372&  	     &  1?&	P1=0.054  &A1=0.01 &  71 &A5me  &B=F7V	       \\	      
V474     Mon&   40535&Q=0.020	     &many& P1=0.1361 &A1=0.15 &  19 &F2IV	  & -	       \\
UZ       Lyn&   43378&  	     &    &	0.065?  &      &  60	 &A2Vs        & -      \\		  
Y        Cam&   +76 286~    &  	     &  1 &	0.0665  &0.04v & 46	 &A9IV        &K1IV	       \\    
UX       Mon&   65607&  	     &    &	 0.04:        & 0.04   & 37 &G2IIIv & -	       \\    
SZ       Lyn&   67390&  	     &  1 &	0.12	& 0.51 & 40 &A9 	      &F2-3	   \\	 
RS       Cha&   75747&  	     &    &	0.077	  & $\leq$0.01& 65 &A8IV	&A8IV	   \\	 
DL       UMa&   82620&  	     &    &	0.0831  &0.06  &  -  &F0	      & -	   \\	 
KZ       Hya&   94033&  	     &    &	0.0595  &0.8   &$<45$& -	      & -	   \\	 
FM       Vir&  110951&N  	     &  1 &	0.0756v &.01-.04&  24 &A8m	   &A7         \\     
V644     Her&  152830&  	     &    &	0.1151  &0.040 &  20 &F5II	  & -	       \\      
$o$      Ser&  160613&	    	     &    &	        &      & 135 &A2Va        & -	    \\ 	   
V577     Oph&+06 3679&               &    &      0.0695	& 0.050  &  -  &A	    & - 	 \\
$\delta$ Sct&  172748&  R+NR	  &$>$4&    P1=0.1938 &A1=0.29 &  30 &F2IIIp& - 	 \\				     
CE       Oct&  188520& P1\ap0.2d	  &    &      P2=0.05451&A2=0.006&  -  &A7IV/V      & - 	 \\			       
18       Vul&  191747&               &    &      0.1215	& 0.008  &  45 &A3III	    & - 	 \\
$\delta$ Del&  197461& R (comp B    )&3$^{*}$&0.1568 & 0.050 &  29 &A7III	    &A7p $\delta$ Del\\		 
IK       Peg&  204188&  		  &    &      0.044   &0.01  &  37 &A8m 	& -	 \\		 
GX       Peg&  213534& Q=0.020;W	  &5&   0.056   &0.015 &  46 &A5m 	& -	 \\					
$\beta$  Cas&     432& Q=0.020	  & 1  &      0.1010  &0.015 &  69 &F2III-IV	& - \\
GN       And&    2628& Q=0.017(R)    &1   &	 0.069   &0.02v?&  18 &A7III	   & -      \\     
AV       Cet&    8511&  	     &    &	 0.0685    &$\leq$0.01& 195 &F0V	 & -	    \\    
V479     Tau&   24550&  	     &    &	 0.0758    & 0.03   &  -  &F3II-III    & -	    \\  	       
V480     Tau&   30780& Q=0.021       &    &	 0.04	 &0.02  & 145 &A7IV-V	   &G3V     \\
OX       Aur&   50018&  	     &    &	 0.11544 &	& 150 &F2V	   &G5V     \\
$\rho$   Pup&   67523&	R	  &1   &      0.14088 &0.09  &  15 &F2mF5IIp	& -	 \\	 
FZ       Vel&   77140& Q=0.016       &    &	 0.065:  & 0.02 &  59 &Am	       & -	    \\    
RX       Sex&   90386&  	     &    &	 0.08	   &$\leq$0.01&  -&A3V         &B=G5	    \\
FG       Vir&  106384&  	     &    &	 0.0786    & 0.040  &  21 &Am	       & -	    \\  			  
$\alpha$ Lyr&  172167&		  &    &        	&	 &   5 &A0Vvar      & - 	 \\	
V1644    Cyg&  192640&		  &    &      0.031	& 0.020  &  25 &A2V &B=G8III	 \struutdown \\	
\hline
\end{tabular}
\vspace{-5mm}
\tablenotetext{}{$^{*}$ More frequencies are present, due to pulsation in comp A, also $\delta$ Scuti}
}
\end{center}
\end{table}

\section{Visual binaries}

\subsection{Orbits and physical parameters}

In Table~\ref{Table:VB} we present additional information regarding the physical parameters 
of the components of some of the visual binaries listed in Tables~1~
and \ref{Tab:variability} Given are the name and HD designation, the orbital period in years, 
the semi-axis major in AU, the sum of the masses, and if known the radii $R_{1}$, $R_{2}$
and the individual masses $M_{1}$, $M_{2}$ as well as a reference.\\
We recall that the sum of the masses, $\Sigma M$ (in solar mass), and the orbital elements, 
P$_{\rm orb}$, the orbital period (expressed in years), and A, the true semi-axis major (in AU), are 
linked through Kepler's third law:
\begin{displaymath} \Sigma M = \frac{{\rm A}^{3}}{{\rm P}^{2}}. \end{displaymath}
There are three VB/O cases ($\delta$ Ser, $\gamma$ CrB, $\tau$ Cyg) but only two have a
reliable orbit determination. Of course, only if the mass ratio is known from either 
the absolute astrometry or the spectroscopy can precise individual masses be obtained.

$\sigma^{2}$ CrB is not listed since it is very probably not a $\delta$ Scuti 
star. This orbital pair is in fact a triple system. The visual binary
has a period of $\approx$ 1000 yr and a semi-axis major of 140 AU.
It also contains a double-lined and chromospherically active spectroscopic binary. 
From differential spectrophotometry, Frasca et al. (1997) deduce that there is 
{\bf no} evidence for a 0.1 day periodicity but that all the variation is linked
with the period of rotation. They conclude that the photometric variability is due 
to dark spots on the secondary component of the SB and very probably 
{\bf not} of $\delta$ Scuti type. This conclusion is also supported by the
spectral classification (F6+G0).
We also discarded the following two visual double stars: V377 Cas (sep=2.1\arcsec) 
because the light variations are not typical of the $\delta$ Scuti type (Lowder 1989)
and DL Dra (HR 5492, sep=3.9\arcsec), shown to be probably constant already
in 1990 (Papar\'o et al. 1990).

We next focus on some particular objects of this category.

\begin{table}
\caption{$\delta$ Scuti stars in visual binaries} \label{Table:VB}
\begin{center} \scriptsize
\begin{tabular}{rrccccccccc}
\multicolumn{2}{c}{Designation} &  P$_{\rm orb}$ &  A  & $\Sigma M$ & $R_{1}$  & $R_{2}$ & $M_{1}$ & $M_{2}$ & Remark &Refs\\
Name                      & HD  &  years         &  AU & $M_\odot$ & $R_\odot$& $R_\odot$&$M_\odot$&$M_\odot$&        &    \\
\hline
VW           Ari&   15165& \ap 4.10$^{5}$& \ap 8700 &  4.0 & 3.5 & 1.6 & 2.2 & 1.8 & & AC95 \\
$\theta^{2}$ Tau&   28319& $<$8.10$^{5}$& \ap 15000 &  $>4.5$  & - & - & 4.5 &  -  & & TSL97 \\ 
KW           Aur&   33959& $<$21000     & \ap 1200 &  $>3.7$  & - & - & 3.7 &  -  & & FW79 \\  
$\kappa^{2}$ Boo&  124675& \ap 8500      & \ap 640  &  3.5     & 2.8 & 1.5 & 2.1 & 1.4 & & FJ95 \\		    
$\delta$     Ser&  138917& 3000:        & 320:     &  3-4:    & - & - & - & -  & B var? & BP89 \\ 						     
$\gamma$     CrB&  140436& 92.7         & 32.5     &  4.2     & - & - & - & -  & e=0.51 & SS99 \\  				       
$\tau$       Cyg&  202444& 49.9         & 18       &  2.5     & - & - & - & -  & e=0.25 & BP89 \\ 					
    -        Phe&  220392& \ap1.10$^{5}$& \ap 3300 &  4.1     & - & - & 2.3 & 1.8 & corr. P & LV99 \\	  
\end{tabular}
\end{center}
\tablenotetext{}{\scriptsize References: AC95 = Abt \& Morrell (1995);
BP89 = Baize \& Petit (1989); FJ95 = Fu \& Jiang, 1995; FW79 = Fitch \& W\'\i sniewski (1979);
LV99 = Lampens \& Van Camp (1999); SS99 = S\"oderhjelm (1999); TSL97 = Torres et al. (1997)}
\end{table}

\subsection{Selected stars}

\subsubsection{VW Ari A} \label{vwari}

This very wide binary consists of an A-type primary showing $\lambda$ Boo peculiarities
in the spectrum and a F-type secondary of solar-like composition in common spatial
motion (similar proper motions (CCDM, Dommanget \& Nys 1995) and radial velocities 
(Fehrenbach et al. 1987)). 
Because of its very wide separation and the different chemical composition of
its components, stellar capture was proposed as the probable mechanism of formation. 
However, Chernyshova et al. (1998) recently argumented that capture is improbable 
and also no longer needed to explain the differences in composition since these could
be due to the specific evolution of the primary star solely. 
The primary is the $\delta$ Scuti star that was intensively observed by the STEPHI 
network in 1993 (Liu et al. 1996). They detected more than seven frequencies,
more or less grouped. Non-radial rotationally split modes might therefore be present 
in this star, a medium to fast rotator with vsin{\it i}=90 kms$^{-1}$. Andrievsky et al. (1995) 
concluded from their spectral analysis that VW Ari A has no sharp lines while the lines 
of VW Ari B are "sharp and strong", so fast rotation is not applicable to the companion. 
Such a difference in rotational velocity between both components was not considered by 
Liu et al. (1996), who also found a much smaller surface gravity value for component A 
via photometric calibration, thereby resulting in masses surprisingly small to match 
the given spectral types. 

\subsubsection{$\theta^{2}$ Tau} \label{the2tau}

$\theta^{2}$ Tau is the most massive main-sequence star of the Hyades cluster,
the primary of a common proper motion pair with $\theta^{1}$ Tau (at a separation of 5$\farcm$6),
and a member of a single-lined spectroscopic binary (SB1) of period 140.7 days with an (highly eccentric) interferometric orbit 
from the Mark III optical interferometer (Pan, Shao, \& Colavita 1992; Hummel \& Armstrong 1992). 
Torres, Stefanik \& Latham (1997; hereafter TSL97) determined individual masses and the 
distance of the system by treating it as a double-lined spectroscopic binary (SB2), thereby  exploring a range of values for the 
mass ratio and the rotational velocity. The derived orbital and the Hipparcos trigonometric 
parallaxes agree very well. This spectroscopic binary is formed by two stars of nearly identical 
colour and mass but with different projected rotational velocities: TSL97 obtained a best fit 
with v$_{B}$sin{\it i}= 110~kms$^{-1}$ while v$_{A}$sin{\it i}= 80~kms$^{-1}$. Both are therefore considered
to be rapid rotators. From the location in a colour-magnitude diagram and a best fit with an 
isochrone of age $\approx$ 630 Myr, they conclude that the primary is in a phase near H core 
exhaustion, immediately preceding the phase of overall contraction. However, because both the 
binarity and the fast rotation may affect the colour indices, its evolutionary status may
still be ambiguous (see also Kr\'olikowska 1992).  \\
Various multi-site campaigns have been conducted. Breger et al. (1989; hereafter BG89) obtained 
five closely spaced and stable frequencies, all of which had amplitudes below 0.01 mag. They 
discarded rotational splitting since it could not explain the observed frequency separations and 
proposed a mixture of modes of different {\it l} and {\it m} values. Kennelly et al. (1996) 
discussed a large set of radial velocity and line profile data from which up to seven 
frequencies emerged with only three frequencies in common with the previous analysis. They suggested 
long-term ($>$ 6 yr) amplitude variability and a combination of low and high degree modes.
Amplitude variability on a 10 yr time scale is also claimed by Li, Zhou \& Yang (1997). 
Both components lie within the $\delta$ Scuti instability strip but it seems well established 
that the more massive primary is the pulsating star (BG89, KW96). Even though a wealth of 
information about physical parameters is known for $\theta^{2}$ Tau, the situation regarding variability
is very confused and there is up to now no clear mode identification possible for the apparently 
very complex (not solved) frequency spectrum. 

\subsubsection{KW Aur A} \label{kwaur}

KW Aur A is an ellipsoidal, single-lined spectroscopic binary system
(Harper 1938) as well as the primary component of the visual multiple
system ADS 3824 that is in common proper motion with the tertiary component
situated at a distance of some 1000 AU with a period of about 24000 yr
(Tokovinin 1997). Though well-detached, both components of the inner
system must be tidally deformed due to the short period of the binary (P=3.79d). 
Photometric and radial velocity variations analyzed together gave evidence for 
three non-radial modes (Fitch \& Wisniewski~1979) (FW79). In addition, they also 
identified a frequency that corresponds to twice the orbital frequency due to the 
ellipticity effects on the mean light curve. They suggested a single pulsation 
frequency split by tidal modulation (instead of rotational modulation) as the cause 
for the observed non-radial triplet (same {\it l}, different {\it m}). In an analysis 
of line profile variations for this star, Smith (1982) corroborated their conclusion.\\
Rotational effects are expected to be small due to the fact that there is synchronization
of the rotation with the orbital motion (see also Table 2). More theoretical work needs 
to be done in this context: computations should be done for inhomogeneous models, with 
a density gradient from core to envelope, better applicable to $\delta$ Scuti stars in 
general and to KW Aur A in particular. 

\subsubsection{$\kappa^{2}$ Boo A} \label{kapboo}
 
$\kappa^{2}$ Boo forms a common proper motion system with $\kappa^{1}$ Boo, 
itself consisting of a spectroscopic binary with a period of 1791~days
and a high eccentricity (Batten, Fletcher, \& MacCarthy 1989). Both stars have 
colours that locate them in or near the $\delta$ Scuti instability strip. 
The intrinsic separation is $\approx$ 600 AU, consequently the orbital period could 
be about 8000 yr long if one adopts 3.5 M$_\odot$ as the sum of the masses.
The primary is a fast rotator but this is improbable for the secondary
($v_{B}$sin{\it i}=40 kms$^{-1}$). 
Frandsen et al. (1995) (FJ95) observed both adopting a scheme of programme 
star (comp. A) versus comparison star (comp. B) in order to study the 
pulsation behaviour of $\kappa^{2}$ Boo. They detected a multiperiodic 
pattern with up to four frequencies of which three are very close and
one is a probable radial mode. All associated amplitudes are below
0.01 mag. FJ95 found a matching model by fitting a common isochrone
through both stars and derive a distance that is in perfect agreement
with the (later published) Hipparcos distance. They are hesitant to
invoke rotationally split modes since their model produces a good match
of the observed set of frequencies, even though rotation was not 
considered! One problem of this analysis is that they were not able to
identify which component is responsible for the variability, let stand
for what frequencies. Since component B is an early F-type binary, in
principle it could also be a pulsating variable star (as admitted by 
the authors themselves). It would thus be very interesting to precisely
identify the source of the multiple frequencies detected in this triple
system.  

\subsubsection{HR 8895} \label{hr8895}

An interesting common origin pair is formed by HD 220392 and HD 220391.
Both components are located in the $\delta$ Scuti instability strip
but up to now only the primary seems to present short-period pulsations 
of the $\delta$ Scuti type. More observational effort should be spent 
on both stars to investigate their behaviour with respect to pulsation.
For a more detailed analysis we refer to Lampens \& Van Camp (1999). 
A very recent and exciting information concerns new radial velocity
measurements: Grenier et al. (1999) confirm that the radial velocities 
are in agreement and furthermore also show that component B has a
variable radial velocity! 

\subsubsection{$\gamma$ CrB} \label{gamcrb}

This is an orbital binary with a period of 93 yr, a high eccentricity and a semi-axis 
major of some 33 AU (Hartkopf \& McAlister 1989). The system consists of a B9/A0 IV 
primary, a suspected Maia star, and a 1.5 mag fainter A3/A4 main-sequence secondary. 
Lehmann et al. (1997) measured the radial velocity of the system and concluded that 
stars of the Maia type can pulsate in the same way as $\delta$ Scuti stars. In a first 
possible scenario they invoke an additional mechanism for the amplitude variation of 
the NR modes, while a second possible scenario implies spontaneous excitation and 
damping of these modes. One more alternative is that component B could well be the 
$\delta$ Scuti variable star.

\subsubsection{DG Leo} \label{dgleo}

  This system is also triple, consisting of a double-lined spectroscopic binary
and a visual component B that is the $\delta$ Scuti pulsating star. The spectroscopic
double has an orbital period of 4.147 days and a mass ratio close to unity. 
All three stars are possibly nearly identical with a global spectral type
A8IV and with marginal metallicity. Rosvick \& Scarfe (1991) (RS91) detected a velocity change
of the visual component with respect to the centre of mass of the spectroscopic pair
that is due to the orbital motion. Speckle measurements show that this system
is highly inclined, possibly also highly eccentric (Hartkopf, McAlister, \& Mason, 1999). 
This was also claimed by RS91, who presented evidence for the occurrence of very shallow eclipses
with a possible inclination of $\approx$~70\deg~. They obtained a mass of some 
2~M$_\odot$ for each spectroscopic component. Resolving the visual orbit will lead
to more accurate masses for all three stars.\\
Only one periodicity seems to be known for the $\delta$ Scuti variable companion 
but more may be present since amplitude and phase changes have been reported (RS91). 
It is a very interesting multiple system where both metallicity and binarity effects 
can potentially influence the pulsational characteristics (on stabilization and tidal mixing
see Budaj (1996; 1997)).

\subsubsection{$\epsilon$ Cephei}
This is the primary of a visual double system that is a chance alignment of
a bright foreground and a much fainter background star ($\rho$=128";
$\Delta {\rm m}$=5). However, $\epsilon$ Cephei was recently also found to be a 
new astrometric binary discovered by the Hipparcos satellite (ESA 1997, 
part DMSA/G). We have no further binarity information.
Observations by Lop\'ez de Coca et al. (1979) in the B band
indicate two periods, perhaps three. Ratios are compatible with the second 
overtone and the fundamental radial modes. But the amplitudes are very 
small. From an analysis of the line profiles, Baade et al. (1993) found evidence 
for intermediate to high order NR p-modes with 6$\leq \mid m \mid \leq$8. Why such high 
order p-modes are also detected in integrated light data is unclear. This star 
shows a rich spectrum of modes according to Horner et al. (1996) and it is 
not improbable that it still belongs to the main sequence. Given its spectral type of F0, 
it is a medium to fast rotator.

\subsubsection{$\beta$ Cas} \label{betcas}

      ADS~107 A was reported to show radial velocity variations with an apparent
period of about 27 days (Mellor 1917) and was therefore subsequently classified as 
SB (for example in $\delta$ Scuti stars catalogues). However Abt (1965) already 
concluded that there was no evidence for such a binarity, as was later on confirmed by 
the results of Yang, Walker, \& Fahlman (1982). They showed that the radial velocity 
curve varied with a very short period, close to the photometric period of pulsation. 
In conclusion, $\beta$ Cas is a monoperiodic $\delta$ Scuti pulsator with a very 
stable small amplitude but for which the mode identification is uncertain (Rodr\'{\i}guez 
et al. 1992; Riboni, Poretti, \& Galli 1994). We have classified it among the 
probable optical visual double stars and do {\bf not} consider it as a pulsating star 
in a binary system (cfr. bottom part of Tables~1~and~2).

\section{Spectroscopic and eclipsing binaries}

An interesting feature of pulsating stars in binary systems is the so-called
{\it light-time effect}: the orbital motion of the variable star produces a
Doppler effect, and the observed period of variation will decrease and increase.
This effect, together with mass transfer in a semi-detached system and apsidal motion, 
can be the cause of regular period changes in a binary system.
For a circular orbit of radius $a$, orbital period $P$ and inclination $i$, the time of light
maximum is given by (e.g. Barnes \& Moffett, 1975):
\begin{equation}
T_{\rm max} = T_o + EP_o - \frac{a\sin i}{c}\cos 2\pi \left(\frac{EP_o}{P}-\phi\right),
\end{equation}
where $E$ is the number of elapsed periods, $T_o$ is the initial epoch of maximum, 
$P_o$ is the pulsation period and $c$ is the speed of light.
The light-time phenomenon has been observed in many binary $\delta$ Scuti stars and offers a useful
way to obtain the orbital period.\\
Another feature of very short-period binaries is tidal deformation: the components
can no longer be considered as spherical objects but have a triaxial symmetry
with the longest axis directed toward the companion. The effects on the pulsation
may be quite diverse and very difficult to detect: a) frequency shifts; b) amplitude 
modulation in the case of radial pulsation (e.g. $\theta$ Tuc, De Mey et al. 1998);
c) frequency splitting in the case of non-radial pulsation (e.g. KW Aur, Fitch \& 
Wi\'sniewski 1979)....

\subsection{Single-lined spectroscopic binaries}

Table \ref{Table:SB1} contains the orbital elements for the $\delta$ Scuti stars which are
reported in the literature as SB1. The table lists the name, HD designation (or BD number 
if HD does not exist), spectral type, orbital period in days, eccentricity, semi-amplitude of the 
radial velocity in kms$^{-1}$, mass function in solar mass and finally the reference of 
the orbit.
For simplicity, when the orbit was published in the Catalogue of Batten et al. (1989), we only refer
them and not the original papers. In this case, the spectral type is also the one they quote.
Let us recall here that the mass function is given by 
\begin{equation}
f(m) = \frac{M_2}{(M_1+M_2)^2} \sin ^3 i,
\end{equation}
where $M_1$ and $M_2$ are the mass of the primary and secondary, respectively,
and $i$ is the unknown inclination of the orbit with respect to the line-of-sight. 

We  do not claim to have the complete list of $\delta$ Scuti members of spectroscopic binaries. The
classification of a $\delta$ Scuti star as a spectroscopic binary is indeed not easy if no detailed
study is done. A simple variation of the radial velocity over the years may not be an indication 
of duplicity but solely the result of pulsation. The misclassification of AI CVn (King \& Liu, 1990)
is very illustrative in this respect.
Szatm\'ary (1990) gave a list of $\delta$ Scuti stars in binary systems, either eclipsing or spectroscopic. 
Some stars of his table do not appear in our Table \ref{Table:SB1} : ET And, DV Aqr, MM Cas, RX Cas, AZ CMi, ZZ Cyg and $\theta$ Vir. \hfill \break
\begin{itemize}
\item The binary system ET And contains a B9p primary which was believed to be pulsating. 
However, Weiss et al. (1998) have
found that it is the main comparison star, HD 219891, which pulsates, with a period of
0.1 day and a semi-amplitude of 2.5 millimag. Thus, HD 219891 is 
the $\delta$ Scuti star while ET And appears very stable!
\item The $\delta$ Scuti nature of DV Aqr needs to be confirmed.
\item MM Cas is an eclipsing binary reported by Chaubey (1983) to have brightness fluctuations of 
0.08 mag amplitude and a period of $3^{\rm h}40$. Chaubey therefore conjectured that the star is
 a $\delta$ Scuti variable. This needs confirmation. 
\item RX Cas is a semi-detached system whose orbital period is increasing. The A-type primary spectrum is
that of a shell or disc that completely conceals from view the primary star.
\item AZ CMi is a pulsating star of spectral type F0 III. The long orbital period (2625 days) 
quoted by Szatm\'ary (1990) 
is derived from a sinusoidal fit to the O-C data, assuming that they can be attributed to the
light-time effect. A spectroscopic study of this object would thus be useful.
\item Frolov et al. (1982) could not confirm the $\delta$ Scuti status of ZZ Cyg. We did not retain it in our list. 
\item Finally, $\theta$ Vir (HR 4963 = HD 114330) also deserves some attention. This star is a spectroscopic binary with 
a 17.84 years orbit, which is itself a visual binary. The star has been
classified as a hot Am star and Beardsley \& Zizka (1977) detected a variability at the level
of 5 kms$^{-1}$ with a period of $0.^{\rm d}152360$. Adelman (1997), however, found no 
evidence for variability, although Scholtz et al. (1998) note that there is "remarkable
scatter which could possibly indicate real variations". These authors also remark that Hipparcos
revealed a variability at the 8 millimag level with a period of $0.^{\rm d}697382$. From their
spectroscopic investigation, they detected a well defined period of $0.^{\rm d}0614$, 
with an amplitude of 0.4 kms$^{-1}$. It could be the case that $\theta$ Vir is another
example of a pulsating Am star in a binary system. This needs however to be more definitively
ascertained. 
\end{itemize}

There are some other stars which could enter our list of $\delta$ Scuti in multiple systems.
Among these pending cases, let us consider the high amplitude $\delta$ Scuti variable,
 V474 Mon (HR 2107) which is quoted in the Yale Bright Star
Catalogue (see King \& Liu 1990) as a $15.^{\rm d}492$ spectroscopic binary, although no trace of any
orbit could been found in the literature. It is also noteworthy that the quoted binary period
is exactly twice the value of the period of the Blazhko-effect (Romanov \& Fedotov 1979).
Another example is V650 Tau (HD 23643), which is quoted by Garcia et al. (1995) as spectroscopic binary.
No mention of this could be found in the literature. Abt et al. (1965) give several radial velocity
measurements in the range -35 to 2 kms$^{-1}$, with errors up to about 10 kms$^{-1}$. Unless a detailed
study is performed, this is however not a proof for binarity. 

CC And is probably a binary with an orbital period of 10.469 days (Fitch, 1976) and 
an eccentricity of 0.12 (Fitch, 1969). The shape of the light curve does indeed vary
with this period, a phenomenon attributed to tidal modulation of the fundamental by a
faint companion. Fitch (1967) detected 6 pulsation frequencies in the light curve, 
while 7 were found more recently by Fu \& Jiang (1995).

\begin{table} 
\caption{\label{Table:SB1}
$\delta$ Scuti stars in single-lined spectroscopic binaries}
\begin{center} \scriptsize
\begin{tabular}{rrlrlrll}
\multicolumn{2}{c}{Designation} & Sp.T. & P$_{\rm orb}$ & e & K          & f(m)      &  refs \\
Name                      & HD  &       & days          &   & kms$^{-1}$ & M$_\odot$ &  \\
\tableline
RZ Cas      & 17138  & A2V   & 1.1952 & 0. & 70.1~ & 0.043& B89 \\
$\rho$ Tau  & 28910  & F0V   & 488.5~~~  & 0.09     & 18.5~ & 0.32~~    & B89\\
V483 Tau    & 27397  &  F0IV  & 2.486~  & 0.03     & 29.9~ & 0.0069&  K99 \\
V775 Tau    & 27628  & A3m    & 2.1433 & 0.       & 26.6~ & 0.0042&   B89\\
$\theta ^2$ Tau & 28319& A7III& 140.728~& 0.75    & 31.0~ & 0.13  &   B89\\
KW Aur      & 33959  & A9IV  & 3.7887 & 0.       & 23.0~ & 0.0048~~&  B89\\
UZ Lyn      & 43378  & A2V   & 20.819~ & 0.367    & 3.77 & 0.000093&  S98\\
Y Cam       & +76 286~& A9IV  & 3.3055 & 0.       & 35~~~   & 0.015 &   B89\\
SZ Lyn      & 67390  & F2     & 1118.1~~~ & 0.188    &  9.6~ & 0.101~~  & M88\\
V644 Her    & 152830 & F3Vs  & 11.8586 & 0.365 & 27.44& 0.02054  & BI82\\
IK Peg      & 204188 & A8m    & 21.724~ & 0.       & 41.5~ & 0.16   & B89 \\
GX Peg      & 213534 & A5m    & 2.34~~   & 0.02:    & 84.9~ & 0.15   & B89\\
\end{tabular}
\end{center}
\vspace{-5mm}
\tablenotetext{}{\scriptsize References: B89=Batten et al. (1989); BI82=Bardin \& Imbert (1982); 
K99=Kaye (1999); M88=Moffett et al. (1988); 
S98=Scholtz et al. (1998)}
\end{table}

We will now further discuss some of the stars listed in Table \ref{Table:SB1}.
\subsubsection{UZ Lyn}

The orbital elements for UZ Lyn (2 Lyn = HD 43378 = HR 2238) 
quoted in Table~\ref{Table:SB1} are only preliminary and need
confirmation. It was obtained by Scholtz et al. (1998), although in their spectroscopic
run covering 240 min, the radial velocity was found to be constant. However,
Caliskan \& Adelman (1997) had published some radial velocity measurements, suggesting the
star to be a spectroscopic binary. Combining their velocities with the one obtained by 
Caliskan \& Adelman (1997), Scholtz et al. (1998) found three orbital solutions with periods
of 21, 33 and 87 days. They kept the 21 days solution as it had the smaller residuals.
Additional data are clearly called for.

\subsubsection{SZ Lyn}

The pulsation behaviour of SZ Lyn was discovered by Hoffmeister (1949). It was later classified as a dwarf Cepheid by 
Broglia (1963) and is now considered as 
a monoperiodic (0.$^{\rm d}$12) high-amplitude $\delta$ Scuti star. 
Moffett et al. (1988) improved the value of the pulsation period to 0.12052115 days. This period is apparently undergoing a secular change of 3~$10^{-12}$ d/cycle (Soliman et al. 1986).
McNamara (1997) derived a semi-empirical $P-L$ relation of SX Phe and large-amplitude 
$\delta$ Scuti stars :
\begin{equation}
<M_v> = -3.725 \log P - 1.933
\end{equation}
For SZ Lyn, he quotes an absolute magnitude of $M_v$=1.35 and a period of
0.1205 d, as well as a mass of 1.92 M$_\odot$ and a radius of 3.18 R$_\odot$.
Rodr\'\i guez et al. (1996) found that, like all other high-amplitude  $\delta$ Scuti and SX Phe variables, it is a radial pulsator. Rodr\'\i guez (1999), 
analyzing all available photometric datasets, did not find any long-term change of amplitude of the light curve.

The binary nature of SZ Lyn was first suggested by Barnes \& Moffett (1975) as an explanation for the periodic ephemeris needed to account for the observed times of maximum. The expected period was around 1146 days. 
The binary nature was confirmed by CORAVEL radial velocity
measurements by Bardin \& Imbert (1981) who obtained a preliminary eccentricity of 0.26. They also obtained a total amplitude in radial velocity of 39.9 kms$^{-1}$, which, combined with their preliminary orbit, 
suggests a variation of 0.115 R$_\odot$ over one pulsation cycle for the radius of the star.
With additional observations, Bardin \& Imbert (1984) however obtained
a slightly longer (1181.5 d) and less eccentric (0.191) orbit. 
Using photometric data, Soliman et al. (1986) found the orbital period to be 1173.5 $\pm$ 2 days. 
Moffett et al. (1988), using both photometric and spectroscopic data,  determined an orbital period of 1181.1 days and an eccentricity of 0.188. The value of the mass function, 0.101 M$_\odot$, implies
that the unseen companion is most likely on the main sequence with a spectral type between F2 and K3, that is, in the mass range 0.7 - 1.6 M$_\odot$. 

\subsubsection{V644 Her}

Like FM Vir (see below), this star was first known as a spectroscopic binary before
being noticed as a variable. The first published orbits gave orbital 
periods of 11.$^{\rm d}$848, 11.$^{\rm d}$857, 11.$^{\rm d}$878 and 11.$^{\rm d}$851. 
Bardin \& Imbert (1982) used CORAVEL to derive more precisely the
orbital elements and obtained a period of 11.858592 days and an eccentricity of 0.365. From Hipparcos data, the absolute magnitude of the system is M$_{\rm v}$ = 1.87, a possible value for a F2~IV star of about 1.5 - 2 M$_\odot$. The absence of the secondary from the spectra as well as from the
CORAVEL trace, implies that the secondary is fainter by at least two magnitudes, corresponding to a star cooler than F6-8~V, and therefore 
less massive than about 1.2 M$_\odot$. 
Breger (1973) showed the variable nature of the star, with a period of 0.098 day and an amplitude of 0.02 mag. 
Elliott (1974) found the period to be 0.1150 day with an amplitude of 0.044 mag.
It might be of interest to note that the ratio of the orbital to the pulsation period is exactly an
integer value, 121 when using Breger's period, and close to 103 when using 
Elliott's one (see also Sect. \ref{Sect:Dis}).

\subsubsection{GX Peg}

From the results of their three weeks multi-site campaign, Michel et al. (1992) unambiguously 
detected five frequencies. 
Goupil et al. (1993) identified  
two radial modes of order n=2,3 and one non radial mode l=1, n=3 split by rotation. \hfill\break
The orbital elements of GX Peg were first determined by Albitzky (1933) and
by Harper (1933). Although the two sets of data were contemporary (data obtained
between 1928 and 1932, and between 1926 and 1933, respectively), there is a
difference of 6 kms$^{-1}$ in the radial velocity amplitude of the derived orbits. 
Bolton \& Geffken (1976) obtained 25 spectrograms betwen 1971 and 1974 to derive a new orbit.
They also recomputed the orbit from Albitzky and Harper data. Although they 
found a good agreement betwen Albitzky's orbit and theirs, the significant 
discrepancy between the periastron longitudes lead them to conclude
to the possibility of apsidal motion with a period of about 260 years. The 
very small eccentricity of the orbit makes this possibility unlikely. 
The Lucy \& Sweeney (1971) test indeed indicates that the eccentricity is compatible with zero. 
As the typical time to have synchronization between orbital and rotation motion is 
smaller than the time needed to circularize an orbit, this 
small orbital period (2.34 days) system has certainly achieved synchronization. The rotational
velocity therefore implies a value of R $\sin i \simeq 2.75 $ R$_\odot$. 
Goupil et al. (1993) concluded that the rotational splitting of the modes are indicative 
of the fact that the
star cannot rotate as a a solid body and is thus not synchronized down to the core. 

\subsection{Double-lined spectroscopic binaries}

Table \ref{Table:SB2} gives the orbital elements of those $\delta$ Scuti stars
which are classified as SB2 and for which we could find details in the literature. 
The outline of the table is similar to Table \ref{Table:SB1} except
that we mention the two spectral types when available, as well as two semi-amplitudes of radial velocity. We also list the mass ratio instead of the
mass function. 

To this list, we should also add BQ Cnc (HD 73729) which has been classified
as an SB2 (Abt \& Biggs, 1972) but for which no orbital elements are known. 
Again, some cases of either misclassification, either lack of precise elements can be
mentioned.
56 Ser (HD 160613) for example, is listed in the Bright Star Catalogue as having two spectra,
while only one spectrum is visible on the Parkins plate (Slettebak, 1954).
On the other hand, DL UMa (HD 82620) is quoted by Henriksson (1979) as an eclipsing binary 
with a period of $0.^{\rm d}42$, while he previously (Henriksson 1977) considered it as a $\delta$ Sct variable with an amplitude of 0.056 mag and a period of $0.^{\rm d}0831$. If confirmed, the almost 
exact integer ratio between the pulsation period and the orbital period might be of great 
interest. 

When studying CE Oct (HD 188520), Kurtz (1980) discovered a second period of $0.^{\rm d}21$ which
may result from a g-mode pulsation or ellipsoidal variability. Morris (1985) rejected the 
latter as being the least likely.

\begin{table}
\caption{$\delta$ Scuti stars in double-lined spectroscopic binaries} \label{Table:SB2}
\begin{center} \scriptsize
\begin{tabular}{rrllrlrrll}
\multicolumn{2}{c}{Designation} & Sp.T$_1$&Sp.T$_2$ & P$_{\rm orb}$ & e & K$_1$&K$_2$  & q & refs \\
Name & HD & & &days & & \multicolumn{2}{c}{kms$^{-1}$} &  & \\
\tableline
AI Hya      &  +00 2259 & F2  & F0  & 8.29~~~ & 0.23 & & & 0.92 & A91 \\
$\theta$ Tuc & 3112 & A7 IV & & 7.1036~ & 0 & 8.57 & 95.6 & 0.0896  & DM98\\
WX Eri      & 21102 & A5 & K0 & 0.82327 & & & & 0.7 &  BD80 \\
UX Mon & 65607 & G2 III & A5 III-IV & 5.9~~~~ & 0. & 60.~~ & 140.~ & 0.44&  B89 \\
RS Cha& 75747 & A8 IV & A8 IV & 1.6699 & 0. & 136.1~ & 138.9 & 0.98 & B89 \\
FM Vir & 110951 & F0 IIIm & & 38.324~~ & 0.074 & 48.05 & 52~~ & 0.92 &  B57\\
18 Vul      & 191747 & A3 III & & 9.316~~  & 0.       & 78.5~ & 86.3 & 0.91     & B89 \\
$\delta$ Del & 197461 & A7 IIIp & & 40.58~~~ & 0.7 & & & & R76 \\
\end{tabular}
\end{center}
\tablenotetext{}{\scriptsize References: A91=Andersen (1991); B57=Bertiau (1957); B89= Batten et al. (1989); BD80= Brancewicz \& Dworak (1980); DM98= De Mey et al. (1998);
R76=Reimers (1976)}
\end{table}

\subsubsection{$\theta$ Tuc}

Cousins \& Lagerwey (1971) were the first to notice the variability of $\theta$ Tuc and to derive a period
of variation around 70-80 minutes. Later, Stobie \& Shobbrook (1976) classified the star as a $\delta$ Sct star and Kurtz (1980) determined a set of 8 stable frequencies. This was later extended to 10 highly-stable frequencies by Papar\`o 
et al (1996). 
From their high-resolution spectra, De Mey et al. (1998) could derive 4 frequencies, the most significant one corresponding to the main photometric
frequency. They showed that this pulsation mode is radial. 
A new frequency, $f_2=18.82$ c/d was found, not appearing in the
photometric data. It must therefore correspond to a high degree pulsation mode. 

Papar\`o et al (1996), noting periodicity of the long-term variations, 
suggested for the first time that $\theta$ Tuc might be member of a binary system.
This was confirmed by Sterken (1997) which showed the star to be in a non-eclipsing binary 
system  with ellipsoidal variations, with a period of 7.$^{\rm d}$04 and a mass ratio of about 0.1-0.15. Both 
the primary and the secondary minima were observable in the light curve. De Mey, Daems \& Sterken 
(1998) made an extensive study of this object. Using high-resolution spectroscopy, they could classify the system as a double-lined spectroscopic binary with a circular orbit and they derived the orbital elements listed in Table~\ref{Table:SB2}. The orbital period (7.1036 d) is not very far from the
previously determined photometric period. The resulting mass ratio is $q=0.0896$, a rather low value for a SB2. De Mey et al. (1998) concluded that
the mass of the components are constrained by $M_1 < 4.3 M_\odot$ and
$M_2 < 0.4 M_\odot$, while the radii of the primary and secondary lie 
between 1.7 $R_\odot$ and 2.6 $R_\odot$, and 1.6 $R_\odot$ and 2.2 $R_\odot$, 
respectively. One can therefore believe that $\theta$ Tuc is a post-mass transfer 
binary, in which the secondary is probably the remnant of an Algol-like 
mass-losing star. 
Sterken (1997) emphasizes the strong similarities between the orbital light curves of $\theta$ Tuc
and HD 96008, a system with a very small mass ratio. It has to be noted however that while in the
case of HD 96008, one of the component nearly fills its Roche lobe, this is not the case for
$\theta$ Tuc. Using De Mey et al. (1998) results, one can see that the primary lies well inside 
its Roche lobe, while the secondary fills maybe only about 50 \% of it.
It is also noteworthy that $\theta$ Tuc seems to show a rotational velocity (v$\sin i$=80 kms$^{-1}$) too large
for synchronization between the orbital and the rotational motion.

\subsubsection{FM Vir}

FM Vir (=32 Vir) is one of the very few Am stars reported as a pulsating star.
FM Vir was recognized as a metallic-line star by Roman, Morgan \& Eggen (1948). 
The Am phenomenon is generally attributed to slow rotation and the effect
of diffusion (e.g. Abt \& Morrell, 1995). Bertiau (1957) quotes 
v$\sin i$ of 20 km$^{-1}$ for FM Vir.\hfill \break
The velocity variation was detected by Adams (1914) and the first orbit was determined by Cannon (1915) who found an orbital period of 38.3 days and also detected double-lines. Petrie (1950) confirmed the double-lined nature of FM Vir and derived a magnitude difference between the two components of 0.43 $\pm$ 0.11 mag. Bertiau (1956) derived a new orbit, with an orbital 
period of 38.3 days, a  rather typical value for Am stars. In the spectral 
region in which he was observing ($\lambda \lambda$ 4350-4650), he could not observe
the lines of the fainter star. However, the lines of the primary are shallow, as if filled in by the continuous spectrum of a companion. He also
observed a fairly large scatter of the individual velocities around the 
mean velocity curve, something which can now be explained by the variability of
the component. \hfill \break
Bartolini, Grilli \& Parmeggiani (1972) found the star FM Vir to be variable with an amplitude ranging from 
$0.^{\rm m}02$ to $0^{\rm m}.05$ and a period of about $0.^{\rm d}07$. Bartolini et al. (1983) confirmed that the star is pulsating with a strongly variable amplitude. The period P=$0.^{\rm d}07188$ has the highest amplitude and is constant.
Kurtz et al. (1976) performed a photometric and spectroscopic study of FM Vir. Their light curves had an observed range in the visual amplitudes between 0.01 and 0.035 mag, while the derived periods for the different nights range from 0.07 to 0.084 days, with an average of $0.^{\rm d}0756$, 
clearly suggesting the $\delta$ Scuti character of the star. The Am character of the star was also confirmed. They also concluded that
FM Vir is a double-lined binary system and obtained v$\sin i $ values of 
24 $\pm $ 6 kms$^{-1}$ and 140  $\pm $ 25 kms$^{-1}$ for the primary and
the secondary, respectively. Therefore, while the primary has a rotation slow
enough for the Am phenomenon to appear, the secondary is above the observed
cutoff for Am stars. The secondary must then have normal abundances. 
The properties of the components place them both inside the instability strip. Because Breger (1970), from an extensive survey, found that classical Am stars do not pulsate, the suggestion was made that it is the secondary which pulsates. Assuming a magnitude difference between the two components of 0.43 mag and a light variability of the system of 0.035 mag indicates that the pulsational amplitude of the secondary should be 0.09 mag.
Mitton and Stickland (1979) suggested that the primary is an evolved Am 
star (or $\delta$ Delphini star) of subgiant luminosity, which pulsates, 
while the secondary is a main-sequence star of spectral type near A7. 
They were able to detect the secondary in their spectra and to derive a magnitude difference between the two components between 0.6 and 0.9 mag. A
semi-amplitude for the secondary could also be determined, leading to 
a mass ratio of 0.92 $\pm$ 0.03. This, combined with the mass function of the system, implies that the system must be nearly edge-on for the masses
not to be unrealistically large. Masses around 2.05 M$_\odot$ and 1.89 M$_\odot$ are derived for the primary and the secondary, respectively.

FM Vir is clearly a system of great interest. The fact that we have two 
components of almost the same mass, but with different chemical composition
should lead to a better understanding of stellar evolution mechanisms. 
Although it is clear that the primary owns its Am status to its slow rotation, one could 
wonder why two very similar stars in a close binary
would have a different rotation. This should certainly not be the case if 
the slow rotation was due to tidal effects. However, the orbital period might be too large for synchronization to have taken place (e.g. Levato 1976). 
The clue may lie in the subgiant status of the primary~: being more evolved, the star's rotation decreased. This should not be much more than a factor 2
however, so that we conclude that there was already a large difference in
rotational velocity between the two components when the system was formed. 

\subsubsection{$\delta$ Del}

The spectroscopic binary nature of $\delta$ Del was discovered by Frost (1924). Eggen (1956) discovered its photometric variability 
with a period of 0.$^{\rm d}$13505 and a variable range in brightness, with a mean around 0.05 mag. Spectroscopic observations by 
Struve, Sahade \& Zebergs (1957) indicated that the radial-velocity is variable with a period of 0.$^{\rm d}$13447. 

$\delta$ Del is the prototype of a sub-class among the $\delta$ Scuti stars with metal-line subgiant or giant spectra which might be characteristic of
evolved Am stars (Breger 1979). Preston (1973, quoted in Reimers 1976) found the star to be an 
eccentric double-lined spectroscopic binary with a 40.5 days orbital period.
Both components are $\delta$ Scuti variables and Reimers (1976) showed that they have identical chemical compositions, i.e. all metals up to the iron group are deficient relative to the Sun by a factor 2, while the abundance of heavy 
elements (Sr, Y, Zr, Ba, Ce, La and Eu) are enhanced by factors between 4 and 8 relative to iron. 
Smith (1982) considered the fact that both components could be $\delta$ Scuti stars
and concluded that comp A is probably a monoperiodic radial pulsator while comp B,
with the strongest $\delta$ Del peculiarity, oscillates in a mixture of radial and
non-radial modes.
Baade et al. (1993) confirmed that $\delta$ Del is a SB2 with well separated
and almost identical spectra.

\subsection{Eclipsing binaries}

Among the stars quoted in Tables \ref{Table:SB1} and \ref{Table:SB2}, some are
eclipsing systems. Such systems are very powerful tools in astrophysics
as they allow the determination of the individual masses and radii of the
components. 

\begin{table}
\caption{$\delta$ Scuti stars in eclipsing binaries} \label{Table:Ecli}
\begin{center} \scriptsize
\begin{tabular}{rlrrrlrrr}
\multicolumn{1}{c}{Name} & Sp.T$_1$& M$_1$ & R$_1$ & Teff$_1$ 
 & Sp.T$_2$& M$_2$ & R$_2$ & Teff$_2$ \\
\tableline
Y Cam	& A9 IV & 1.9~ & 3.15 & 
	& K1 IV & 0.4~ & 3.05 & \\
AB Cas  & A3 V  & 1.78 & 2.9~ & 8000
	& K1 V  & 0.39 & 1.7~ & 4460\\
RS Cha  & A5V   & 1.86 & 2.14 & 
        &       &      &      & \\
WX Eri  & A5    & 2.23 & 1.84 & 7400
        & K0V   & 1.56 & 1.73 & 5840\\
AI Hya 	& F2    & 2.15 & 3.92 & 6700 
	& F0    & 1.98 & 2.77 & 7100 \\
UX Mon  & A6V   & 3.47 & 4.30 & 
        &       &      &      & \\
\end{tabular}
\end{center}
\vspace{-5mm}
\end{table}

Here again, some misleading examples appeared in the literature. For example, 
RY Lep is among the list of southern eclipsing binaries observed by Popper (1966) but he
could only observe sharp single lines.
Diethelm (1985) showed that this star is actually a relatively bright high-amplitude $\delta$
Scuti star with a period of $0.^{\rm d}2254$ and an amplitude of 0.35 mag.

\subsubsection{AB Cas}

As reviewed by Rodr\'\i guez et al. (1998), AB Cas is one of the clear examples where the 
light curves simultaneously and clearly show both types of variability: binarity and 
pulsation. It is an Algol-type binary system with a period of $1.^{\rm d}3668$, while
the primary component is a monoperiodic $\delta$ Sct star with a pulsation period of $0.^{\rm d}0583$. 
The star is pulsating in the fundamental radial mode. 
Rodr\'\i guez et al. (1998) using a previously determined value for the mass ratio of 0.22 and
assuming the secondary to fill the Roche lobe, found the parameters quoted in Table~\ref{Table:Ecli}.

\subsubsection{Y Cam}

Together with AB Cas this is one of the oldest $\delta$ Sct variables known to be in an
eclipsing system: its variability was discovered in 1903 by Mrs. Ceraski. 
Its components (A9 IV and K1 IV) have rather different masses (1.9 M$_\odot$ and 0.4 M$_\odot$)
but rather similar radii (3.15 R$_\odot$ and 3.05 R$_\odot$, respectively). The light curve 
is very reminiscent of the Algol-type and the
system is thought to be semi-detached, with the K star filling its Roche lobe. 
Shapley (1917), using the data of Miss Harwood, computed the first orbital elements of Y Cam, although a light curve was already obtained by Nijland as
quoted in the catalogue of Shapley (1913).
Dugan (1924) obtained a complete visual light curve and computed a solution 
for the system. It was shown that the dimensions of the two components 
were nearly the same. Szcepanowska (1955), on the basis of 176 times of minimum, 
found it necessary to introduce a sinusoidal term into the elements of Y Cam which
Plavec et al. (1961) thought to be only the main part of the whole period 
variation. They moreover concluded that the large periodic term could not
be due to a third body nor to apsidal motion. 
Y Cam nevertheless appeared in the catalogue of systems with apsidal motions
of Petrova \& Orlov (1999), with an apsidal period of 60 years.
Broglia \& Marin (1974) revealed the primary to be a $\delta$ Sct star with a period of  $0.^{\rm d}0634697$ and a variable light amplitude. The amplitude variation is not correlated with the phase in the orbital motion which led them to conclude that the "amplitude variations in $\delta$ Sct stars are not necessarily caused by a companion". 
Broglia \& Marin (1974) confirmed the variation of the orbital period, which they considered
as a proof for mass transfer in the system.

\section{Discussion} \label{Sect:Dis}

Does the binarity influence the pulsation properties of $\delta$ Scuti stars?
In the light of the previous sections, it is obvious that there is no clear
and easy answer. Every discussed case seems to be particular in its own way.
Apart from the binarity, many other phenomena are involved that may even
have a stronger impact on the pulsation properties of the stars: evolution,
chemical composition, rotational effects...  
In an effort to generalize, some authors consider the ratio between the 
pulsation period and the orbital period in order to search whether a resonance 
mechanism between both can occur (e.g. Frolov et al. 1980, Tsvetkov \& Petrova 1993). 
This is however only possible if both periods are determined with very high accuracy. 
For example, in the case of SZ Lyn, the ratio between both periods is almost exactly 
9800 (taking the orbital period from Moffett et al. (1988)). But with a slightly 
different value of the orbital period (1181.5 instead of 1181.1), this ratio, 9803.26,
is not indicative. Such an exercise is therefore only applicable to very close systems 
for which tidal interactions are expected to be important! 
We computed the ratio between the orbital and pulsation periods for the
closest binary systems of our list (with orbital periods up to 20 days) and, 
in general, we do not find values close to an integer. Five systems however form an exception: 
KW Aur, CC And, DL Uma, V644 Her and WX Eri. DL Uma and WX Eri have very short 
orbital periods (0.42 and 0.82 days, respectively) and in both cases, this is 
almost exactly 5 times the pulsation period. For KW Aur, CC And and V644 Her, 
this ratio is 43, 84 and 103, respectively. It is not clear whether these last values 
have any physical meaning at all.\\ 

In very close systems tidal forces may force the system towards synchronization. 
For spectral types typical of $\delta$ Scuti stars, this happens for binary systems 
with orbital periods of a few days up to 10 days if the stars are evolved. 
Synchronization will slow down the star. Thus, an indirect effect of binarity may be 
the change of the rotational velocity. Possible correlations between
the amplitude of the pulsation and the rotation have been discussed
before (e.g. Breger 1980; Solano \& Fernley 1997). Such a correlation is also seen
in the data presented here.\\

Some cases present better evidence than others: e.g. amplitude modulation in the case 
of the radial pulsator $\theta$ Tuc (De Mey et al. 1998) or possible frequency splitting 
in the case of the non-radial pulsator KW Aur (Fitch \& Wi\'sniewski 1979) but other effects
may also occur (e.g. frequency shifts) that are almost impossible to detect.
We need better models to check whether any of the above presented assumptions 
on the link between binarity and pulsation are valid and whether these can explain
the observational facts in the most interesting cases.

Nevertheless, we should carefully investigate $\delta$ Scuti stars 
in binary systems from the observational point-of-view since they provide
additional constraints on the physical parameters of the pulsating star,
and therefore also on the characteristics of the pulsation. For single
field $\delta$ Scuti stars, the position in the HR diagram and thus the
evolutionary phase may be ambiguous, especially at the end of the core
H burning phase, while $\delta$ Scuti stars in binary systems can be located
with much better accuracy. It may also be especially worthwhile to study
the {\it differences} in variability between two nearly identical components 
of a binary system, of which one or both may be $\delta$ Scuti stars. 
Various such cases have appeared in the present discussions and many 
deserve further observations and study. 

\bigskip

\acknowledgements This research made use of the Simbad database, operated at CDS, Strasbourg, France, as 
well as of the NASA Astrophysics Data System. We also acknowledge funding by project G.00265.97 of the 
{\it Fonds voor Wetenschappelijk Onderzoek (FWO, Belgium)}. 


\end{document}